\documentclass[12pt,preprint]{aastex}
\usepackage{graphics}

\newcommand{\FUSE}{{\it FUSE}} \newcommand{\EUVE}{{\it EUVE}}
\newcommand{\Chandra}{{\it Chandra\/}}

\begin{document}

\title{An Assessment of the \ion{Fe}{18} and \ion{Fe}{19} Line Ratios
from the \Chandra\ Grating Observations of Capella}

\author{P. Desai, N. S. Brickhouse, J. J. Drake, A. K. Dupree,
R. J. Edgar, R.~Hoogerwerf, V.~Kashyap, and B.~J. Wargelin}
\affil{Harvard-Smithsonian Center for Astrophysics, 60 Garden Street,
Cambridge, MA 02138} \email{pdesai@cfa.harvard.edu,
nbrickhouse@cfa.harvard.edu, jdrake@cfa.harvard.edu,
adupree@cfa.harvard.edu, redgar@cfa.harvard.edu,
rhoogerwerf@cfa.harvard.edu, vkashyap@cfa.harvard.edu,
bwargelin@cfa.harvard.edu} \author{R. K. Smith} \affil{NASA Goddard
Space Flight Center/JHU, Code 662, Greenbelt, MD 20771}
\email{rsmith@milkyway.gsfc.nasa.gov} \author{D. P. Huenemoerder}
\affil{MIT Center for Space Research, 70 Vassar Street, Cambridge, MA
02139} \email{dph@space.mit.edu} \author{D. A. Liedahl}
\affil{Lawrence Livermore National Laboratory, P.O. Box 808, L-41,
Livermore, CA 94551} \email{liedahl1@llnl.gov}

\doublespace

\begin{abstract}

Observations of \ion{Fe}{18} and \ion{Fe}{19} X-ray, EUV, and FUV line
emission, formed at the peak of Capella's ($\alpha$ Aurigae) emission
measure distribution and ubiquitous in spectra of many cool stars and
galaxies, provide a unique opportunity to test the robustness of \ion{Fe}{18}
and \ion{Fe}{19} spectral models. The Astrophysical Plasma Emission
Code (APEC) is used to identify over 35 lines from these two ions
alone, and to compare model predictions with spectra obtained with the
\Chandra\ Low Energy Transmission Grating and High Energy Transmission
Grating Spectrometers, the {\it Far Ultraviolet Spectroscopic
Explorer}, and the {\it Extreme Ultraviolet Explorer}. Some flux
discrepancies larger than factors of two are found between
observations of \ion{Fe}{18} and \ion{Fe}{19} lines and predictions by
APEC and other models in common usage. In particular the X-ray
resonance lines for both ions are stronger than predicted by all
models relative to the EUV resonance lines. The multiwavelength
observations demonstrate the importance of including dielectronic
recombination and proton impact excitation, and of using accurate
wavelengths in spectral codes. These ions provide important diagnostic
tools for $10^7$~K plasmas currently observed with \Chandra, {\it
XMM-Newton\/}, and \FUSE.

\end{abstract}


\section{Introduction} \label{intro}
 
The Capella system (HD 34029, $\alpha$ Aurigae), consisting
principally of two cool giant stars (G8 III + G1 III), is one of the
strongest coronal X-ray sources, and offers an opportunity to
benchmark the models used in the interpretation of X-ray spectra from
astrophysical plasmas. Although plasma codes have gone through major
improvements (e.g., Mewe et al. 1995; Brickhouse et al. 1995; Smith et
al. 2001; Young et al. 2003), their accuracy and completeness for
diagnostic analysis at high spectral resolution has yet to be fully
assessed. Brickhouse \& Drake (2000) proposed the use of the \Chandra\
grating calibration observations for a series of studies, collectively
known as the Emission Line Project (ELP), to test spectral models of
collisionally ionized plasmas. This paper compares current spectral
models for \ion{Fe}{18} and \ion{Fe}{19} with \Chandra\ observations
of Capella as a step towards ensuring that astrophysical
interpretations of spectra are based on a sound understanding of the
physical processes involved. The Capella spectrum is well studied,
showing no evidence for flares (Brinkman et al. 2001; Canizares et
al. 2000; Brickhouse et al. 2000). The emission measure distribution
(EMD) of the Capella system shows a strong narrow peak at 6~MK, near
the temperature of peak emissivity for \ion{Fe}{18} and \ion{Fe}{19},
producing numerous strong transitions. Furthermore, line fluxes from
these ions show only modest variability ($\sim$20\%) over timescales
of months to years (Dupree et al. 2005), validating the combined
analysis of multiple observations with \Chandra.

Although Fe L-shell (i.e. \ion{Fe}{17} to \ion{Fe}{24}) X-ray lines
offer powerful diagnostic potential for collisionally ionized plasmas,
two long-standing atomic modeling problems originate from solar
observations of neon-like \ion{Fe}{17} (see Saba et al. 1999) that
have only recently been addressed by laboratory programs (Brown et
al. 1998; Laming et al. 2000; Beiersdorfer et al. 2002). Similar
discrepancies between models and observations of other ions are now
arising from \Chandra\ spectra. For example, Xu et al. (2002) find
that the observed \ion{Fe}{18} $3s$-$2p/3d$-$2p$ ratio in the
elliptical galaxy NGC~4636 is higher than predicted by APEC and similar
to the ratio observed in Capella. The analogous \ion{Fe}{17} ratio
shows the same pattern of discrepancy, suggesting a common atomic
physics origin. The ELP observations of Capella with three instruments
offer a unique opportunity to compare models and observations over a
broad spectral range. Comparisons among FUV, EUV, and X-ray lines of
\ion{Fe}{18} and \ion{Fe}{19} are particularly useful since the strong
n=2$\rightarrow$2 lines are essentially entirely produced by direct
collisional excitation, and thus should be easier to interpret than
FUV or X-ray lines, which can include contributions from other
processes, such as proton impact excitation and dielectronic recombination (DR).

\section{Spectral Models} \label{spec}

We use the Astrophysical Plasma Emission Code version 1.3 (APEC, Smith
et al. 2001) to predict the Capella spectrum.\footnote{APEC V1.3
models, calculated at the low density limit ($N_e = 1.0$ cm$^{-3}$),
and the atomic rate data used to produce them are available at
http://cxc.harvard.edu/atomdb/.} The APEC models for \ion{Fe}{18} and
\ion{Fe}{19} contain 501 and 994 fine-structure levels, respectively,
up to principal quantum number $n=5$. They include the effective
collision strengths and atomic transition probabilities calculated
using the Hebrew University Lawrence Livermore Atomic Code (HULLAC,
Liedahl et al. 1995). For \ion{Fe}{18}, the collision strengths for
the $2p^5~^2P_{1/2}$ -- $2p^5~^ 2P_{3/2}$ transition include resonance
excitation from R-matrix calculations (Berrington et al. 1998). Proton
impact excitation rates within the ground state are included for
\ion{Fe}{18} (Foster et al. 1994) and \ion{Fe}{19} (R. Reid, private
communication, 1999). Laboratory X-ray wavelengths (Brown et al. 2002)
have been incorporated. APEC currently includes DR rates to excited
levels of \ion{Fe}{17} and H- and He-like ions, but not for the other
Fe L-shell ions. Similarly, DR satellite lines are present in APEC for
\ion{Fe}{17} (Safranova et al 2001), but not for \ion{Fe}{18} and
\ion{Fe}{19}.

\section{Data Analysis} \label{anal}

\subsection{Observations and Data Reduction}
 
Multiple spectra of Capella, acquired between 1999 August and 2002
October include pointings with the \Chandra\ High Energy Transmission
Grating (HETG) with the ACIS-S detector for a total exposure time of
182.2 ks, and with the Low Energy Transmission Grating (LETG) and
HRC-S detector for a total exposure time of 234.2 ks. The HETG and
LETG data, obtained from the \Chandra\ archive, were reprocessed using
CIAO version 3.0\footnote{http://cxc.harvard.edu/ciao/} with only
minor deviations from the standard pipeline procedures. Effective
areas were generated for each dataset using the \Chandra\ calibration
database CALDB 2.8 and were exposure-time weighted to create average
effective areas for the summed spectra.

{\it Extreme Ultraviolet Explorer} (\EUVE) spectra obtained in 1999
September, which are nearly simultaneous with a \Chandra\ LETG/HRC-S
pointing, were processed using standard \EUVE\ Guest Observer software
(IRAF). Agreement between LETG and \EUVE\ fluxes for the lines
discussed in this paper is good to within about 5\%, and henceforth
LETG fluxes will be used. The {\it Far Ultraviolet Spectroscopic
Explorer} (\FUSE) line fluxes are taken from spectra of Young et
al. (2001).
 
\subsection{Modeling and Measurements}

 We calculate the global continuum spectra produced by bremsstrahlung,
radiative recombination continuum, and two-photon emission over the
observed \Chandra\ spectral range. We then fit the temperature of the
continuum model to the line-free regions of the HETG spectrum,
identified both from the APEC line list and by visual inspection,
which yields a temperature of 6~MK, near the peak of the EMD. Since
the LETG spectrum is contaminated by high order emission, the same
continuum model derived from the HETG data is also applied to the LETG
fitting. We adopted the abundances of Brickhouse et al. (2000), who
found no evidence for deviation from the solar abundances of Anders \&
Grevesse (1989). Individual line fluxes from the \Chandra\ spectra were
measured using {\it Sherpa} (Freeman et al. 2001) to fit functions
approximating the instrumental line profiles. Plus and minus orders
were fit separately, with the requirement that the line fluxes be the
same. A narrow range for the FWHM was allowed (for HETG,
0.01 - 0.0135~\AA, and for LETG, 0.045 - 0.06~\AA), standard binning
was maintained, and the Cash statistic was applied (Cash 1979). Table 1
gives the observed fluxes for the \ion{Fe}{18} and \ion{Fe}{19} lines
with $1\sigma$ errors.

\subsection{Model Assumptions} \label{uncertainty}

A continuous EMD (Brickhouse et al. 2000), which is needed to estimate
the contribution of line blends from ions over the entire temperature
range, is normalized to the flux of the \ion{Fe}{18} $\lambda$93.92
resonance line and used to predict the line fluxes given in Table 1.
We note that there is only a few percent difference between the single
temperature continuum model and the EMD for the lines of
interest. Since some \ion{Fe}{19} line emissivities show modest
density-sensitivity between the low density limit and densities
expected under coronal conditions, we have used the APEC code to
compute models for a wide range of densities. The most affected line
ratio is that of $\lambda$101.55 to $\lambda$108.37. At $N_e =
10^{10}$ cm$^{-3}$ the predicted ratio is 0.347 (in photon units) compared with 0.261 at
the standard APEC low density limit, in better agreement with the
observed flux ratio of 0.328.

Lack of significant variability further supports the assumption that
the plasma conditions are stable, as individual lines of \ion{Fe}{17},
\ion{Fe}{18}, and \ion{Fe}{19} show modest flux changes ($< 10$\%
deviation from the average value) between \Chandra\ pointings and the
lightcurves show low levels of variability ($\lesssim$8\%) during a
single pointing. There is also no evidence to challenge the standard
assumptions of negligible optical depth (Canizares et al. 2000; Brown
et al. 1998).

\section{Results and Discussion} \label{disc}

Figure~1 compares the observed \ion{Fe}{18} and \ion{Fe}{19} line
fluxes to those predicted by the spectral codes, APEC, CHIANTI V4.2
(Dere et al. 1997; Young et al. 2003), and SPEX V1.1 (Kaastra et
al. 1996), which incorporates the MEKAL model (Mewe et al
1995). Emissivities provided by M. F. Gu (private communication, 2004)
using the Flexible Atomic Code (FAC, Gu 2003)
are also compared. The fluxes are scaled by the fluxes of their respective strong
EUV resonance lines, for which direct excitation dominates. All models
in the figure are calculated at a single temperature $T_e=6$~MK, and
the same density $N_e=10^{10}$ cm$^{-3}$, except for SPEX, which is at
the zero density limit.

Most striking is the discrepancy between the EUV and X-ray lines: the
observed X-ray fluxes are stronger than predicted fluxes in all
models. Even the X-ray $3d$-$2p$ resonance lines, \ion{Fe}{18}
$\lambda$14.208 and \ion{Fe}{19} $\lambda13.518$, are under-predicted
relative to their EUV counterparts by more than 30\% and a factor of
2, respectively. Since these factors are larger than expected from
calibration errors or line blending, it is possible that the accuracy
of the direct excitation rate coefficients might explain the predicted
weakness of $\lambda$14.208 (see Brown et al. 2005); however, it is
difficult to reconcile that with the larger discrepancy for
$\lambda$13.518.

The \ion{Fe}{18} and \ion{Fe}{19} FUV forbidden line fluxes are in
good agreement with the EUV line fluxes of $\lambda$93.92 and
$\lambda$108.37 for the APEC models, and somewhat better than for the FAC
rates. FAC does not calculate proton impact excitation rates, which
are included in both APEC and CHIANTI. In APEC
models, proton impact excitation increases the forbidden line
emissivities by 15\% and 8\% for $\lambda$974.86 and
$\lambda$1118.07, respectively. The predicted FUV line fluxes also
begin to increase with density above $N_e \sim 10^{12}$
cm$^{-3}$. APEC models give the best agreement at $N_e = 2 \times
10^{12}$ cm$^{-3}$, but are also consistent within observational
errors with the lower coronal density range.

Figure~1 also shows some large discrepancies among the strongest X-ray
lines, reflecting the $3s$-$2p$/$3d$-$2p$ pattern. For these transitions the largest difference
among the predictions results from the number of processes calculated with
each model. For example, even though APEC and CHIANTI have similar
collision strengths for the \ion{Fe}{18} $\lambda$15.625 line,
additional line flux in APEC is produced by direct excitation to $n=4$
and $5$ levels, followed by radiative cascades, while CHIANTI
currently includes only levels only up to $n= 3$. On the other hand,
for \ion{Fe}{18} $\lambda$16.07 APEC and CHIANTI both show differences
of more than a factor of 2 from FAC because neither includes the
effects of DR on the upper level population, which are included in
FAC.

Comparisons of APEC and FAC predictions to the observed fluxes of the
X-ray lines listed in Table~1 are shown in Figure~2. We confirm a general $3s$-$2p/3d$-$2p$ discrepancy pattern for
APEC models that is largely removed with the FAC calculations. The $3s$-$2p$/$3d$-$2p$
ratios of the summed line fluxes from APEC are smaller than the
observed ratios by $\sim$20\%, whereas FAC agreement is within
10\%. The inclusion of DR in the FAC models produces the additional $3s$-$2p$
line emissivity.

Another significant disagreement between the models and observations
occurs for radiative transitions that terminate on excited levels, namely
\ion{Fe}{18} $\lambda$15.870, $\lambda$16.159, and $\lambda$17.623 and
\ion{Fe}{19} $\lambda$15.198 and $\lambda$16.110. Although the APEC
line list, which is reasonably complete in this spectral region, does
not include DR satellite lines from either \ion{Fe}{18} or
\ion{Fe}{19}, blending with satellite lines or lines from other ions
cannot explain the extent of the under-prediction. It is possible that
the large theoretical wavelength inaccuracies for these lines, up to a
few percent, have led to misidentifications in the laboratory
measurements. Blending of nearby lines from the same ion could produce
such a pattern of under and overprediction. For \ion{Fe}{18}
$\lambda$15.870, this latter explanation is consistent with new
wavelength calculations (Kotochigova et al. 2005; M. F. Gu 2005).

\section{Conclusions} \label{conc}

 A surprising result of this benchmark spectral modeling study is the
large discrepancy between modern theory and the Capella observations for the
X-ray and EUV resonance lines of \ion{Fe}{18} (30\%) and \ion{Fe}{19}
(factor of 2). New FAC calculations including dielectronic
recombination bring most X-ray lines into good agreement with
observations; however puzzling
discrepancies as large as a factor of 2 still remain for some
relatively strong lines. Additional laboratory and theoretical work is
needed to eliminate the largest remaining problems. Meanwhile, errors
can largely be minimized by judicious choice of line diagnostics and
consideration of appropriate atomic processes.

\acknowledgements

This work is supported in part by the \Chandra\ X-ray Observatory
Center (NAS8-39073). We thank the CXC staff, particularly Harvey
Tananbaum, for supporting efforts to obtain these data. We also
acknowledge the efforts of the developers of the other public spectral
modeling codes SPEX and CHIANTI as well as M. F. Gu for the atomic
structure code FAC.

\begin{figure}
\plotone{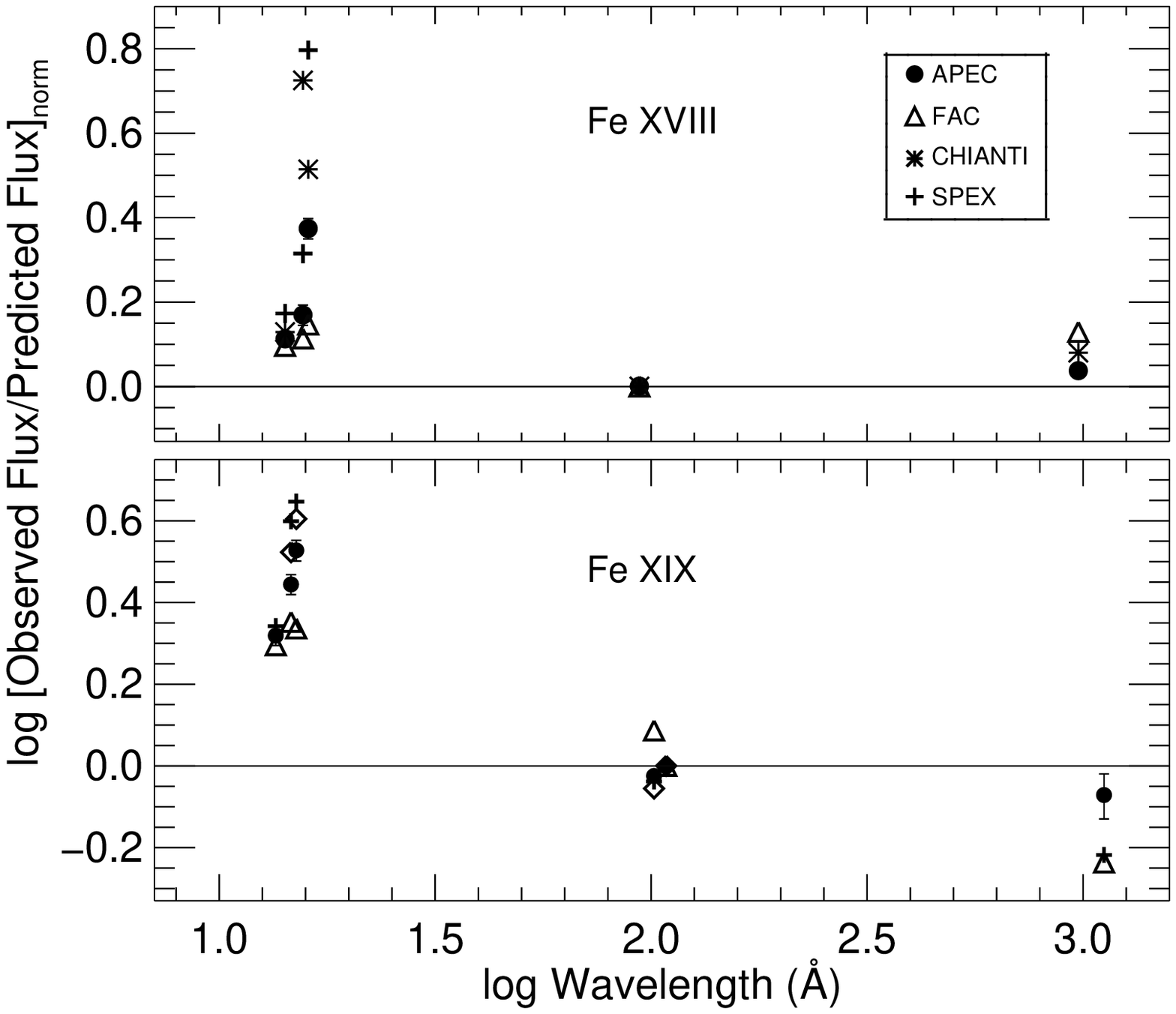}
\vspace{4mm}
\caption{Observed to predicted flux ratios of strong lines in the
X-ray, EUV, and FUV spectral regions. Shown for comparison are the
ratios obtained using the APEC, CHIANTI, and SPEX spectral codes and
the FAC rates. The density is $N_e = 10.0$ cm$^{-3}$, except for SPEX. {\it Top:}
Comparison for \ion{Fe}{18} lines, normalized to $\lambda$93.92. The
X-ray lines plotted here are $\lambda$14.208, $\lambda$15.625, and
$\lambda$16.071. {\it Bottom:} Comparison for \ion{Fe}{19} lines,
normalized to $\lambda$108.37. The X-ray lines plotted are
$\lambda$13.518, $\lambda$14.664, and $\lambda$15.079.}
\end{figure}

\begin{figure}
\plotone{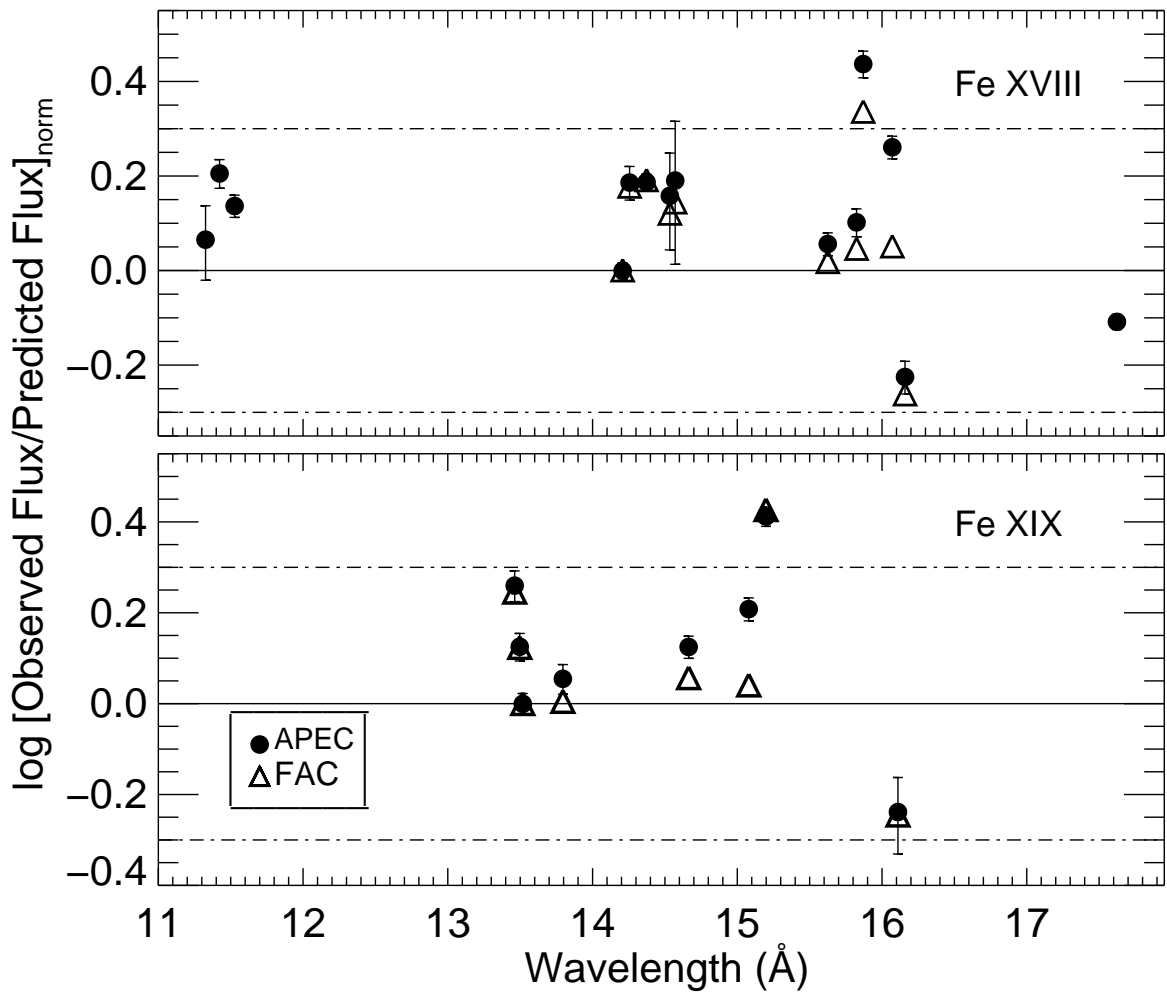}
\vspace{10mm}
\caption{Observed to predicted flux ratios of X-ray lines using FAC
 and APEC. Lines from Table~1 excluding heavily blended \ion{Fe}{18}
 $\lambda$16.004 are shown. Note the $3d$-$2p$ lines are between 14
 and 15 \AA\ for \ion{Fe}{18} and shortward of 14 \AA\ for
 \ion{Fe}{19}. Ratios are calculated at $N_e = 10.0$
 cm$^{-3}$. Dash-dotted lines represent agreement within a factor of
 2. {\it Top:} Comparison for \ion{Fe}{18}, normalized to
 $\lambda$14.208. There are no published FAC models for \ion{Fe}{18}
 $4d$-$2p$ lines around 11.4 \AA. {\it Bottom:} Comparison for
 \ion{Fe}{19}, normalized to $\lambda$13.518.}
\end{figure}
\begin{deluxetable}{ccccr@{ - }lccc}
\tabletypesize{\tiny}

\tablecolumns{8}
\tablewidth{0pt}
\tablecaption{Fe XVIII and Fe XIX Line Measurements}
\tablehead{
\colhead{Inst} & 
\colhead{Ion} &
\colhead{$\lambda_{ref}$} &
\colhead{$\lambda_{obs}$} &
\multicolumn{2}{c}{Transition} &
\colhead{$J_{U}$ -- $J_{L}$}&
\colhead{Model Flux\tablenotemark{a}} & 
\colhead{Observed Flux} \\
\colhead{} &
\colhead{} &
\colhead{(\AA)} &
\colhead{(\AA)} &
\colhead{} &
\colhead{} &
\colhead{} &
\colhead{( ph cm$^{-2}$ ks$^{-1})$} &
\colhead{( ph cm$^{-2}$ ks$^{-1})$} 
}

\startdata

 FUSE\tablenotemark{b} & Fe XVIII &974.86  & 974.85 & $2p^5~^2P_{1/2}$ & $2p^5~^2P_{3/2}$    & $\frac{1}{2}$ -- $\frac{3}{2}$ & 5.063 & $5.50 \pm 0.03 $ \\

 LETG &Fe XVIII  & 103.93 & 103.98 & $2s2p^6~^2S_{1/2}$ & $2p^5~^2P_{1/2}$  & $\frac{1}{2}$ -- $\frac{1}{2}$ &   1.625 & $1.69\pm 0.05 $\\

 LETG & Fe XVIII & 93.923  & 94.02 & $2s2p^6~^2S_{1/2}$ & $2p^5~^2P_{3/2}$    & $\frac{1}{2}$ -- $\frac{3}{2}$ & 4.441  & $4.44 \pm  0.03 $ \\

 MEG  & Fe XVIII & 17.623 & 17.620 & $2p^43p~^2P_{3/2}$ & $2s2p^6~^2S_{1/2}$ & $\frac{3}{2}$ -- $\frac{1}{2}$  &  0.300  & $ 0.30 \pm 0.01$ \\

  MEG  & Fe XVIII &16.159 & 16.163 & $ 2s2p^53s~^2P_{3/2}$ & $2s2p^6~^2S_{1/2}$  & $\frac{3}{2}$ -- $\frac{1}{2}$ &  0.164  & $ 0.13\pm 0.00$ \\

 MEG\tablenotemark{c}   &Fe XVIII & 16.071 & 16.073 & $2p^4(^3P)3s~^4P_{5/2}$ & $2p^5~^2P_{3/2}$   & $\frac{5}{2}$ -- $\frac{3}{2}$ & 0.418  & $0.82 \pm 0.01$ \\

 HEG\tablenotemark{c}  & Fe XVIII & 16.071 & 16.076 & $2p^4(^3P)3s~^4P_{5/2}$ & $2p^5~^2P_{3/2}$   & $\frac{5}{2}$ -- $\frac{3}{2}$ & 0.418  & $1.00 \pm 0.06$  \\

HEG & Fe XVIII\tablenotemark{d} & 16.004& 16.008 & $2p^4(^3P)3s~^2P_{3/2}$ & $2p^5~^2P_{3/2}$   & $\frac{3}{2}$ -- $\frac{3}{2}$ & 0.768   & $0.81\pm 0.04$ \\

 HEG  & Fe XVIII &15.870 & 15.873& $2p^4(^1D)3s~^2D_{3/2}$ & $2p^5~^2P_{1/2}$ & $\frac{3}{2}$ -- $\frac{1}{2}$ & 0.095 & $0.34 \pm 0.02$ \\

 HEG  &Fe XVIII & 15.824  & 15.831& $2p^4(^3P)3s~^4P_{3/2}$ & $2p^5~^2P_{3/2}$  & $\frac{3}{2}$ -- $\frac{3}{2}$ & 0.179 & $0.29 \pm 0.02$ \\

 HEG  & Fe XVIII  &15.625  & 15.628& $2p^4(^1D)3s~^2D_{5/2}$ & $2p^5~^2P_{3/2}$  & $\frac{5}{2}$ -- $\frac{3}{2}$ & 0.290  & $0.43 \pm 0.02$ \\

 HEG  & Fe XVIII&  14.571 & 14.559& $2p^4(^3P)3d~^4P_{3/2}$ & $2p^5~^2P_{3/2}$ & $\frac{3}{2}$ -- $\frac{3}{2}$ &  0.110 & $0.21 \pm 0.07$ \\

 HEG  &Fe XVIII  & 14.534& 14.539& $2p^4(^3P)3d~^2F_{5/2}$ & $2p^5~^2P_{3/2}$ & $\frac{5}{2}$ -- $\frac{3}{2}$ &  0.210 & $0.39 \pm 0.09$\\

 HEG &Fe XVIII  & 14.373  & 14.376&  $2p^4(^3P)3d~^2D_{5/2}$ & $2p^5~^2P_{3/2}$  & $\frac{5}{2}$ -- $\frac{3}{2}$ & 0.278 & $0.55 \pm 0.02$ \\  

 HEG  & Fe XVIII\tablenotemark{e}  & 14.256 & 14.261& $2p^4(^1D)3d~^2S_{1/2}$ & $2p^5~^2P_{3/2}$  & $\frac{1}{2}$ -- $\frac{3}{2}$ & 0.087  & $0.42 \pm  0.03 $\\

\nodata  & \nodata  & \nodata& \nodata& $2p_{1/2}2p_{3/2}^33d_{5/2}$ & $2p^5~^2P_{3/2}$  & $\frac{5}{2}$ -- $\frac{3}{2}$ & 0.141 &  \nodata \\

 HEG  & Fe XVIII  & 14.208& 14.208& $2p_{1/2}2p_{3/2}^33d_{5/2}$ & $2p^5~^2P_{3/2}$   & $\frac{3}{2}$ -- $\frac{3}{2}$ & 0.381 & $1.40 \pm  0.05 $ \\

\nodata   & \nodata    & \nodata & \nodata & $2p^4(^1D)3d~^2D_{5/2}$ & $2p^5~^2P_{3/2}$   & $\frac{5}{2}$ -- $\frac{3}{2}$ & 0.695  & \nodata \\

 HEG  & Fe XVIII   & 11.527 & 11.528& $2p_{1/2}^22p_{3/2}^24d_{5/2}$ & $2p^5~^2P_{3/2}$ & $\frac{5}{2}$ -- $ \frac{3}{2} $ & 0.032 & $0.17 \pm 0.01$ \\

\nodata   & \nodata  & \nodata & \nodata & $2p^4(^3P)4d~^2D_{5/2}$ & $2p^5~^2P_{3/2}$ & $\frac{5}{2}$ -- $ \frac{3}{2}$ & 0.061 & \nodata \\

HEG  & Fe XVIII   & 11.423  & 11.424& $2p^4(^3P)4d~^2F_{5/2}$ & $2p^5~^2P_{3/2}$ &  $\frac{5}{2}$ -- $\frac{3}{2}$ & 0.080 & $0.13 \pm 0.01$ \\

 \nodata & Fe XXII & 11.427 & \nodata & $2s2p_{1/2}3p_{3/2}$ & $2p~^2P_{1/2}$  & $\frac{3}{2}$ -- $\frac{1}{2}$ & 0.007  & \nodata  \\

 HEG  & Fe XVIII   & 11.326  & 11.327& $2p^4(^1D)4d~^2S_{1/2}$ & $2p^5~^2P_{3/2}$ & $\frac{1}{2}$ -- $\frac{3}{2}$ & 0.019 & $0.13 \pm 0.02 $ \\

\nodata   & \nodata  & \nodata  & \nodata & $2p^4(^1D)4d~^2P_{3/2}$ & $2p^5~^2P_{3/2}$ & $\frac{3}{2}$ -- $\frac{3}{2}$ & 0.031 & \nodata \\

\nodata   & \nodata  & \nodata  & \nodata& $2p^4(^1D)4d~^2D_{5/2}$ & $2p^5~^2P_{3/2}$ & $\frac{5}{2}$ -- $\frac{3}{2}$& 0.038 & \nodata  \\

FUSE &Fe XIX\tablenotemark{b,f}  & 1118.07 & \nodata & $2p^4~^3P_{1}$ & $2p^4~^3P_{2}$  & 1 -- 2 & 1.833  & $1.74\pm 0.22$ \\

LETG  & Fe XIX  & 120.00 & 120.04& $2s2p^5~^3P_{2}$ & $2p^4~^3P_{1}$ & 2 -- 1& 0.836   & $0.97 \pm 0.03 $ \\

LETG &Fe XIX  & 111.70 & 111.74& $2s2p^5~^3P_{1}$ & $2p^4~^3P_{1}$ & 1 -- 1 & 0.326  & $0.46 \pm 0.02$ \\

LETG & Fe XIX  & 109.97 & 109.99& $2s2p^5~^3P_{1}$ & $2p^4~^3P_{0}$ & 1 -- 0&  0.413   & $0.46 \pm 0.02$ \\

LETG & Fe XIX  & 108.37 &  108.39 & $ 2s2p^5~^3P_{2}$ & $2p^4~^3P_{2}$ & 2 -- 2 & 3.091   & $ 3.13 \pm 0.05$  \\

LETG & Fe XIX  & 101.55 & 101.59 & $ 2s2p^5~^3P_{1}$ &  $2p^4~^3P_{2}$ & 1 -- 2& 0.838  & $1.02 \pm 0.03$ \\

LETG & Fe XIX  & 91.02  &91.054&  $2s2p^5~^1P_{1} $ & $2p^4~^1D_{2}$ & 1 -- 2& 0.241   & $ 0.45 \pm 0.02 $ \\ 

 HEG & Fe XIX &  16.110 & 16.111&  $2p_{1/2}2p_{3/2}^23p_{1/2}$ & $2s2p^5~^3P_{2}$ &2 -- 2& 0.120  & $ 0.14 \pm 0.03$ \\

 HEG &Fe XIX & 15.198 & 15.204 & $2p_{1/2}^22p_{3/2}^23s$ & $2s2p^5~^3P_{2}$ & 2 -- 2 & 0.080 & $0.39 \pm 0.02$\\

 HEG & Fe XIX & 15.079 & 15.083 &  $2p^3(^4S)3s~^5S_{2}$ & $2p^4~^3P_{2}$ & 2 -- 2 & 0.094 & $ 0.33 \pm 0.02$ \\

 HEG & Fe XIX & 14.664 &  14.671& $2p^3(^2D)3s~^3D_{3}$ & $2p^4~^3P_{2}$ &3 -- 2 & 0.079 & $0.21 \pm 0.01$ \\

HEG & Fe XIX & 13.795 &  13.795& $ 2p_{1/2}2p_{3/2}^23d_{5/2}$ & $2p^4~^3P_{2}$  & 3 -- 2& 0.105 &  $0.24 \pm 0.02$ \\

\nodata & \nodata & \nodata & \nodata&  $2p^3(^2D)3d~^3P_{2} $ & $2p^4~^3P_{2}$  &3 -- 2 & 0.012  &   \nodata \\

 HEG  & Fe XIX  &13.518 & 13.523 & $2p^3(^2D)3d~^3D_{3} $ & $2p^4~^3P_{2}$ &3 -- 2 & 0.262  & $ 0.52 \pm 0.03$   \\

 HEG & Fe XIX &13.497 & 13.507& $ 2p_{1/2}2p_{3/2}^23d_{3/2}$ & $2p^4~^3P_{2}$  & 2 -- 2& 0.118   & $0.32 \pm 0.02$   \\

 \nodata  & Fe XXI &13.507 & 13.507& $ 1s^22s2p_{1/2}^23s$ & $1s^22s2p^3~^3D_{1}$  & 2 -- 2& 0.025   & \nodata   \\

 HEG  & Fe XIX & 13.462  & 13.470 & $ 2p^3(^2D)3d~^3S_{1} $ & $2p^4~^3P_{2}$  & 1 -- 2& 0.072  & $ 0.25 \pm 0.02$ \\

 HEG  & Ne IX & 13.447  & 13.446 & $ 1s^2~^1S_{0} $ &  $1s2p^1P_{1}$  & 1 -- 2& 0.397  & $ 0.40 \pm 0.02$ \\

\enddata

\tablenotetext{a}{\phantom{.}Line blends are listed separately if they contribute $>$ 10\%
to the Fe line of interest (in the model). Fluxes (including blends)
normalized to the $\lambda$93.92 line predicted by the
emission measure distribution model using APEC at a density of 1
$cm^{-3}$ are listed. The observed fluxes have
been corrected for interstellar absorption using $N_H = 1.7 \times
10^{18}$ cm$^{-2}$ (Piskunov 1997), neutral helium, and H/He abundance
ratio set at 11.6 (Kimble et al. 1993). The largest correction at
$\lambda$120.0 amounts to only 9\%.}
\tablenotetext{b}{\phantom{.}See Young et al.\ (2001).}
\tablenotetext{c}{\phantom{.}MEG and HEG measurements of this line are given to
show the cross-calibration. HEG is preferred for this analysis because
of its better spectral resolution.} 
\tablenotetext{d}{\phantom{.}Contribution of OVIII to this line is more than
50\%.}
\tablenotetext{e}{\phantom{.}LETG flux was measured to cross check
  the calibration of LETG vs HETG. The LETG line is somewhat blended,
  but the flux is within 30\% of the HETG flux.}
\tablenotetext{f}{\phantom{.}This FUSE measurement is uncertain as this line is blended. Solar network spectra were used to estimate the contribution of \ion{C}{1} to the blend.}

\end{deluxetable}
\end{document}